# A Tutorial on Network Security: Attacks and Controls

Natarajan Meghanathan
Associate Professor of Computer Science
Jackson State University
Jackson, MS 39217, USA
Phone: 1-601-979-3661; Fax: 1-601-979-2478
E-mail: natarajan.meghanathan@jsums.edu

**Abstract**

With the phenomenal growth in the Internet, network security has become an integral part of computer and information security. In order to come up with measures that make networks more secure, it is important to learn about the vulnerabilities that could exist in a computer network and then have an understanding of the typical attacks that have been carried out in such networks. The first half of this paper will expose the readers to the classical network attacks that have exploited the typical vulnerabilities of computer networks in the past and solutions that have been adopted since then to prevent or reduce the chances of some of these attacks. The second half of the paper will expose the readers to the different network security controls including the network architecture, protocols, standards and software/ hardware tools that have been adopted in modern day computer networks.

## 1. Introduction to Computer Networks

With the phenomenal growth in the Internet, network security has become an integral part of computer and information security. Network security comprises of the measures adopted to protect the resources and integrity of a computer network. This section reviews the basics of computer networks and Internet in order to lay a strong foundation for the reader to understand the rest of this paper on network security.

### 1.1 ISO-OSI Reference Model

The communication problem in computer networks can be defined as the task of transferring data entered by an application user in one system to an application user in another system through one or more intermediate networks [1]. The communication problem is solved using a layered approach through a collection of protocols forming the so-called protocol suite. Each layer, dealing with a particular aspect of the communication problem, is implemented with a particular protocol and the protocols co-operate with each other to solve the entire communication problem. The Open Systems Interconnection (OSI) model [2] is an abstract representation of the basic layers (as stated below and also shown in Figure 1, in top to bottom order) involved to solve the communication problem: Application, Presentation, Session, Transport, Network, Data-link and Physical layers.

The application layer specifies how one particular application uses a network and contacts the application program running on a remote machine. The presentation layer deals with the translation and/or representation of data at the two end hosts of the communication. The session layer is responsible for establishing a communication session with a remote system and it also handles security issues like password authentication before the application user can connect to the remote system. The transport layer provides end-to-end, reliable or best-effort, in-order data packet delivery along with support for flow control and congestion control. The network layer deals with forwarding data packets from the source to the destination nodes of the communication. The data-link layer deals with the organization of data into frames and provides reliable data delivery over the physical medium. The physical layer provides the encoding/decoding schemes and the modulation/demodulation schemes for the actual transmission of data, over the physical medium, as a sequence of bits of 1s and 0s.

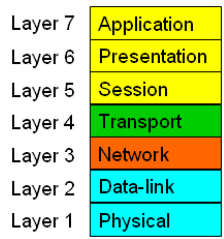
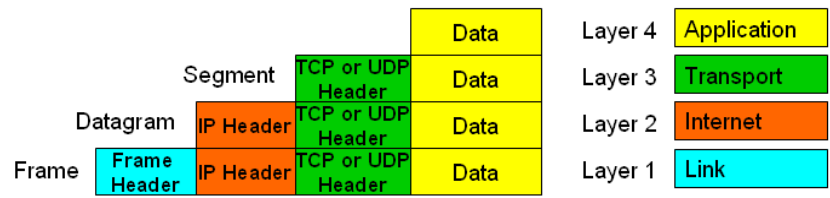

Figure 1: OSI Model          Figure 2: TCP/IP Protocol Stack and the Structure of a Data Packet

## 1.2 TCP/IP Protocol Stack

The seven-layer OSI model is conceptual: it shows the different activities required for communication between application programs running in two different hosts. Its full implementation will result in excessive overhead and will lead to huge delays in data delivery at the destination [1]. The TCP/IP (Transmission Control Protocol/ Internet Protocol) protocol stack [3], shown in Figure 2, is the commonly used model for wide area communications, like the Internet. The TCP/IP protocol stack is composed of the Application, Transport, Internet and the Link layers (from top to bottom). The application layer of the TCP/IP model is in-charge of the responsibilities of the application, presentation and session layers of the OSI model. The transport layer of the TCP/IP model is similar to the transport layer of the OSI model. The Internet layer takes care of addressing and routing the data packets across different heterogeneous networks. Each machine and router in the Internet has a unique IP address. The link layer of the TCP/IP model combines the functionalities of the data-link layer and physical layer of the OSI model. The link layer supports the organization of data into frames and their encoding/decoding mechanisms. The structure and transmission of the frames depends on the topology and hardware technology (like Ethernet, Token Ring and etc) used for the network. A data packet is referred to as segment, datagram and frame at the transport, internet and the link layers respectively.

## 1.3 TCP Connection Establishment

The two commonly used transport layer protocols in the TCP/IP protocol stack are the Transmission Control Protocol (TCP) [3] and the User Datagram Protocol (UDP) [3]. TCP is a connection-oriented, byte-stream based protocol and provides reliable, in-order data delivery. UDP is a connectionless, message-based protocol and provides only best-effort service for end-to-end data delivery. Processes running TCP have to establish a connection before exchanging any data packet. During this connection establishment mechanism, the two processes exchange information about the capabilities and resources available at their respective hosts for the particular communication session that is about to begin. This will help the TCP process running in one host to adjust its data sending rate according to the resources (like the memory buffer space) available for the TCP process at the receiving host. In order to avoid replay errors, the two processes pick an arbitrary starting sequence number for the data packets sent by them. Each byte of data is given a unique, monotonically increasing sequence number. The sequence number of a data packet sent using TCP represents the sequence number of the first byte of the data transmitted in that packet.

    The TCP connection-establishment process (shown in Figure 3) is a three-way handshake mechanism [1] and is explained as follows through this example: Let a process running in host A initiate a session with a process at host B by sending a Synchronization (SYN) packet to host B with the initial sequence number set to X. The process at host A will include information about the memory resources available

(through the 'Advertised Window' field of the TCP header) in the SYN packet. If the process at host B is willing to establish a communication session with the process at host A, then it sends back a SYN/ACK packet that will indicate the memory resources available at host B for this communication, the starting sequence number of the data packets coming from the process at host B and an acknowledgment for receiving the SYN packet from the process at host A. The process at host A will respond back with an ACK packet if it accepts to the advertised window value of host B and is willing to tune down its data sending rate accordingly. Note that the acknowledgment sent to a process/host for receiving a packet with a particular sequence number (say X) indicates the sequence number (X+1) of the next packet expected from the process/host. Typically, host A could be a client and host B could be a server.

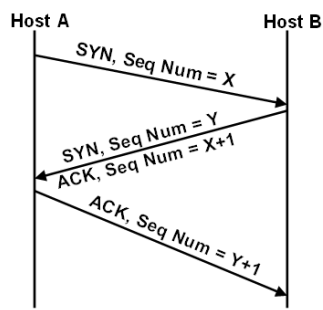

Figure 3: TCP Connection Establishment Mechanism

## 1.4 Internet Control Message Protocol (ICMP)

IP provides best-effort service in delivering datagrams from one host to another host through one or more intermediate networks. The TCP/IP protocol suite provides an error-reporting protocol called the Internet Control Message Protocol (ICMP) that operates in tandem with IP. IP uses ICMP to report errors and certain critical information to the end hosts. Each ICMP message is identified by an 8-bit type field in the IP header. One of the commonly used ICMP message is ECHO Request/Reply [4]. An ECHO request message is sent to the ICMP process running on a host computer to check whether the host is alive. If the host is alive, the host sends a response using the ECHO Reply message.

## 2 Classical Network Attacks

In this section, we describe some of the classical attacks that have exploited the typical vulnerabilities of computer networks and the solutions deployed to combat or reduce the chances of some of these attacks.

## 2.1 Threats in Transit

The network interface card (NIC) [5] of each host in a network is uniquely identified with a hardware address. The NIC will be programmed to pick up only the packets addressed to: (i) The unicast hardware address corresponding to the host, (ii) The multicast hardware address corresponding to the multicast group in which the host is a member of and (iii) The broadcast hardware address. A capable intruder can reprogram the NIC with the hardware address of another host and accept packets addressed to that host. To avoid being caught, the intruder can put a copy of the packet back to the network.

    Wiretapping [6] is the process of extracting information as it flows through a wire. The process of wiretapping differs depending on the communication medium used. In cables, wiretapping can be done through the use of a packet sniffer or through inductance. A packet sniffer [7] is a computer software or hardware that can intercept the traffic passing through a local area network (LAN) cable. A packet sniffer can be used for both beneficial and malicious purposes: (i) To analyze network problems and monitor network usage, (ii) To filter suspect content from network traffic, (iii) To study the structure of the packet

headers of the different protocols used over the network, (iv) To detect network intrusion attempts and (v) To gather information for effecting a network intrusion. As an ordinary wire emits radiation during the propagation of electrical signals through it, an intruder can tap the wire and read radiated signals through inductance without making physical contact with the cable. An intruder intercepting the signals on a broadband cable has to separate the targeted signal from all the multiplexed signals.

Wireless signals are broadcast through the open space and are more susceptible for tapping. For example, the signal path of microwave signals has to be fairly wide to make sure the antenna of the receiver will be hit by the transmitted signal. But, the wider the signal path, the more it is easy for an intruder to interfere with the line of sight of transmission between the sender and the receiver and also to pick up the entire transmission from an antenna located closely to the receiver. Similarly, with satellite communication, there is a tradeoff between coverage and secure communication. A footprint [6] is defined as the pattern produced on the surface of the earth from the satellite's transmitter. A broader footprint is needed to maximize coverage because the signals can be picked up over a huge region. On the other hand, a smaller footprint is desirable to reduce the risk of interception. The angle of dispersion of a satellite transponder is a parameter that could be controlled to adjust the spread of a footprint.

An optical fiber, made of thin glass strands, can carry light pulses over long distances without being much affected by electrical interference [6]. Optical fibers are more secure than any other transmission media because of the following two reasons: (i) Optical fibers are fine tuned to achieve total internal reflection. So, the entire network should be retuned to facilitate tapping and interception and (ii) Optical fibers carry light energy and not electrical signals. So, inductance based tapping would not be possible.

## 2.2 TCP Session Hijacking

TCP session hijacking [8] refers to the act of taking over an already established TCP session and injecting packets into the stream that are processed by the receiver as if the packets are coming from the authentic owner of the session. A TCP session is identified by the quadruple: client IP address, client port number, server IP address and server port number. Any packet that reaches either machine with the above identifiers is considered to be part of the existing session. If attackers can spoof these items, they can pass TCP packets to the client or server and have those packets processed as coming from the other machine.

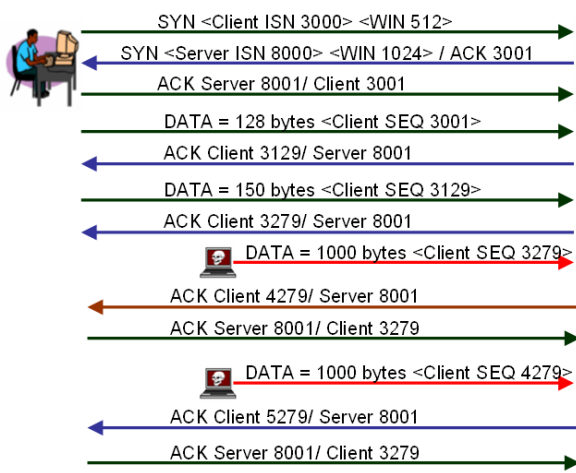
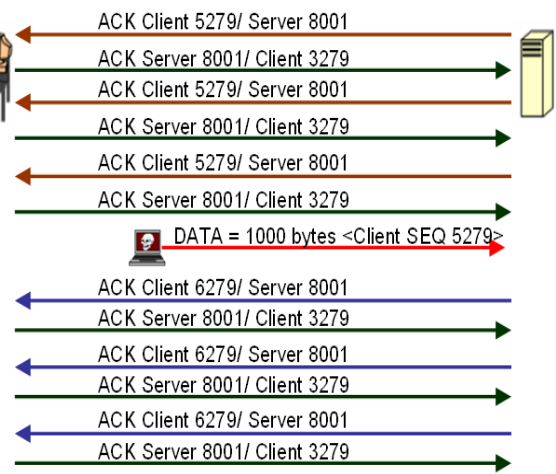

Figure 4: Desynchronizing a TCP Session          Figure 5: Creating a TCP ACK Storm

To successfully hijack an existing TCP session, an attacker has to first desynchronize the session and then inject the intended commands. To desynchronize an existing TCP session (refer Figure 4) between a client and server, the attacker has to first predict the sequence number that is about to be used by a client

(or server) and use that sequence number before the client (or server) gets a chance to use. If the attacker has access to the network, a packet sniffer can be used to look into the packets belonging to the TCP session and one can accurately predict the expected sequence number from the ACK packets exchanged. If the attacker cannot sniff the TCP session between the client and server, then the attacker has to try all possible options and guess the expected sequence number. When the attacker successfully hijacks the TCP session and injects own spoofed data packets (as if the data packets are coming from the original client), the server will acknowledge the receipt of the data packet to the original client by sending it an ACK packet. As this ACK packet will most likely bear a sequence number that is not expected by it, the original client will attempt to resynchronize with the server by sending it an ACK packet with the sequence number that it is expecting. This ACK packet will in turn contain a sequence number that the server is not expecting and so the server will resend its last ACK packet. This cycle will continue and the rapid passing back and forth of the ACK packets creates the TCP ACK storm (refer Figure 5). As the attacker injects more and more data packets, the size of the ACK storm increases and can quickly bring down performance of the network. After a certain number of unsuccessful resynchronization attempts, the original client eventually gets exhausted and closes the connection with the server.

### 2.3 Man in the Middle Attack

With a Man-In-The-Middle (MITM) attack [8], an attacker can read, modify and insert messages between two communicating parties, without either party knowing that the link between them has been compromised. To successfully carry out this attack, one must be able to observe and intercept messages between the two victims. We now describe an example for an MITM attack on public-key cryptography.

Let A and B be the two communicating parties and let M be the attacker who wants to deliver a false message to B. To get started, B sends its public key to A. If M can intercept the communication channel between A and B, then M gets access to the public key of B. Then, M sends A, a spoofed message that claims to have come from B. In this message, M sends its own public key, but A thinks it has received the public key of B. When A sends a data packet to B, it encrypts the packet with (what A considers as) the public key of B and inserts the encrypted message in the channel. M intercepts the message and decrypts it with its own private key to extract the actual message sent by A to B. M then encrypts the message with the public key of B. Note that M could even modify the message before encrypting it again. M inserts the new encrypted message back in the channel so that the message can go to B. B decrypts the message using its own private key and reads the message assuming it came from A.

### 2.4 Echo-Chargen Attack

Chargen (Character Generator) [9] is a protocol of the TCP/IP protocol stack and is used for testing and performance measurement purposes. Chargen runs on TCP port 19 and also on UDP port 19. When a client opens a TCP connection with a server on TCP port 19, the server starts sending arbitrary characters back to the client, until the TCP connection is closed. Whenever a host sends a UDP message to a server on UDP port 19, the server responds back with an arbitrary message and the number of characters in the message will be in the range [0…512].

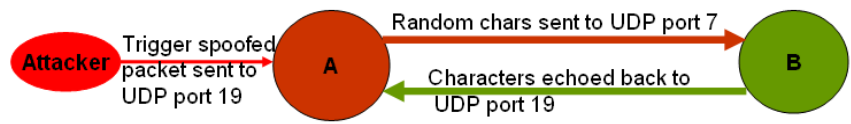

Figure 6: A Typical Echo-Chargen Attack

An attacker can trigger the Echo-Chargen attack by spoofing a conversation between the Echo Request/Reply service and the Chargen service and then redirecting the output of each service to the other, creating a rapidly expanding spiral of traffic in the network. In Figure 6, we see an attacker triggering the attack by sending a spoofed message to one of the targeted hosts (host A) running the Chargen service at UDP port 19. The message is spoofed in such a way that it appears to have originated from the other targeted host (host B) and UDP port 7, which is the port number used for Echo-Request/Reply messaging. Host A now sends a UDP message from port 19 to port 7 of host B. Host B will consider this as an Echo Request message and sends back a Reply message to UDP port 19 of host A. Host A will treat the Reply message as a message received for the Chargen service and sends back a new arbitrary UDP message to port 7 of host B. This cycle of message exchange between the two services will continue and generate excessive traffic in the network. Eventually, the attack consumes memory and processor power at the two targeted hosts A and B and causes them to become non-responsive to user commands.

### 2.5 Smurf Attack

A perpetrator can launch the Smurf attack [8] by sending a spoofed Echo-Request message to a network's broadcast IP address. The spoofed Echo-Request message has the victim's IP address as the source IP address. Hence, each host receiving the broadcast Echo-Request message will send an Echo-Reply message to the victim. The victim will be overwhelmed with a flood of Echo-Reply messages. Thus, the Smurf attack is a kind of Denial-of-Service (DoS) attack. Two solutions have been currently adopted in the Internet to prevent a Smurf attack [10]: (i) Routers do not forward datagrams having the destination address as a broadcast IP address and (ii) Hosts are configured not to reply for Echo-Request messages that were received as a broadcast message.

### 2.6 Traffic Redirection

A compromised router can send out route update messages to all its neighboring routers informing them that it lies on the shortest path to every network in the Internet [11]. The neighboring routers forward all of their incoming data packets to this compromised router, which will get eventually flooded with the data packets and starts dropping them. The data packets do not make it to the destination.

### 2.7 Attacks on Domain Name Service (DNS)

A DNS server is a machine that holds a table (called the DNS cache) mapping the domain names to IP addresses [12]. The server queries other DNS servers higher up in the domain name hierarchy to resolve domain names for which it does not have an IP address entry in its DNS cache and updates its cache with the mapping learnt. DNS cache poisoning [13] is an attack using which the DNS server is made to believe a domain name-IP address mapping as authentic, while, in reality, it is not. Once the DNS cache is poisoned, the entry stays for a while in the cache and affects the clients who use the DNS server in the mean time. For example, an attacker can replace the IP address information for a target file server with the IP address of a compromised file server which the attacker controls. The attacker creates fake entries in the compromised server with file names matching those on the target server. These files could contain malicious contents such as a worm or virus. Users who want to download files from the target file server may end up unknowingly downloading files with malicious content from the compromised file server.

### 2.8   Distributed Denial of Service (DDoS) Attacks

DDoS attacks [8] involve breaking into hundreds or thousands of machines all over the Internet. The attacker installs malicious software on all these compromised machines (called zombies) and controls them to launch coordinated attacks on victim sites. DDoS attacks are normally aimed at exhausting the network bandwidth, overwhelming a router's processing capacity and breaking network connectivity to the victims. The attacker uses any convenient method (like exploiting the buffer overflow attack [6] or tricking the victim to open and install an unknown code from an email attachment) to plant a Trojan Horse [6] on a target machine and transform it into a zombie by also installing a rootkit software. The rootkit helps to conceal the presence of the Trojan Horse and hide its malicious activities. After forming sufficient number of zombies, the attacker sends a signal to all the zombies to launch the DDoS attack on a chosen victim machine. Each zombie may launch the same or a different type of attack on the victim.

### 2.9 Syn Flood Attack

During the TCP connection establishment process, the server maintains a SYN_RECV queue to keep track of the connection requests for which it has allocated the resources and responded back with a SYN/ACK message, but the corresponding ACK from the client has not yet been received. The server eventually times out waiting for the ACK packet and removes the incomplete connection request from its queue. An attacker can launch a DDOS attack by sending several SYN connection request messages using spoofed non-existing IP addresses and never respond back with the ACK messages [8]. The SYN_RECV queue of the server gets filled up with incomplete connection request messages. Even though these incomplete connection requests are discarded after the timeout, if a genuine client attempts to establish a TCP connection with the server in the mean time, the server discards the SYN request from that client.

## 3    Network Security Controls

This section describes several network security controls that have been adopted in modern day computer networks to combat the threats and prevent or reduce the chances of an attack.

### 3.1 Link Encryption Vs End-to-End Encryption

Encryption applied between every pair of hosts connected by a link is called link-to-link encryption [6]. Link encryption is preferred when all the hosts in the network are secure, but the communication medium is shared among several users and is not secure. Almost all the components of a data frame (except the source and destination hardware addresses in the frame header) are encrypted before the frame is inserted onto the physical communications link. As the frame reaches the next hop receiver (could be a router or the end host), the frame is decrypted at the bottom protocol layer and sent to the higher layers for further processing and forwarding. Since encryption is at the bottom protocol layer, the message is exposed in plaintext at all the other layers of the sender and receiver and at the link and Internet layers of the intermediate hosts for hardware addressing and routing. Thus, link encryption protects the message in transit between two computers, but the message is in plaintext inside the end hosts and the intermediate hosts. One or more of the intermediate hosts may not be credible.

Table 1: Comparison of Link Encryption and End-to-End Encryption

| Link Encryption | End-to-End Encryption |
|---|---|
| End hosts of every link should share a key and should be able to do encryption and decryption | The intermediate hosts of a transmission path do not need to have cryptographic facilities. |
| If there are $N$ hosts and $n$ users in a network ($N \ll n$), the number of keys needed would be $N(N-1)/2$ | The number of keys needed for symmetric encryption and public-key encryption would be $n(n-1)/2$ and $2n$ respectively. |
| All message transmissions have to be encrypted and decrypted at every link. | Encryption is application and message specific and need not be done for all messages. |
| One encryption algorithm may be used for all users in all links | Each application user can deploy an encryption algorithm of choice. |
| Data is exposed at the end hosts and the intermediate hosts | Except the application layer, data is encrypted at both the end hosts and the intermediate hosts |

Encryption applied between two application programs running at the end hosts of a communication is called end-to-end encryption [6]. Here, only the data portion of the packet is encrypted at the highest level (i.e. the application layer) and the packet is transmitted with the data in encrypted form throughout the Internet. Thus, end-to-end encryption protects the data against disclosure while in transit, but the data packet could go through potentially insecure intermediate hosts. Table 1 [6] compares the pros and cons of link encryption and end-to-end encryption.

### 3.2 Virtual Private Networks

There are two types of IP addresses: public and private. A public IP address [1] is globally unique and only one machine connected to the public Internet can have a public IP address. Private IP addresses are one of the solutions to reduce the exhaustion of IP address space [1]. A private IP address has to be unique only within the set of networks of a particular organization. Larger organizations have sites at different locations in the world. The hosts in the different sites of the organization may be identified with a unique private IP address. But the same set of private IP addresses can be used in the networks of different organizations. Hence, a packet with a private IP address as the destination IP address cannot be used to route packets from one site to another site of an organization through the public Internet.

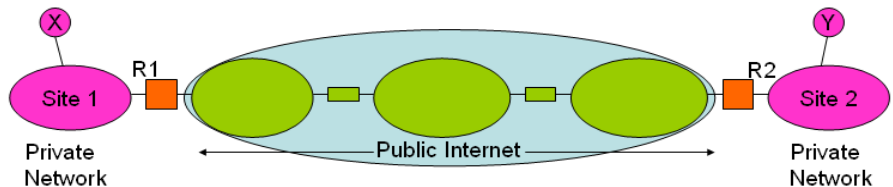

Figure 7: Virtual Private Network

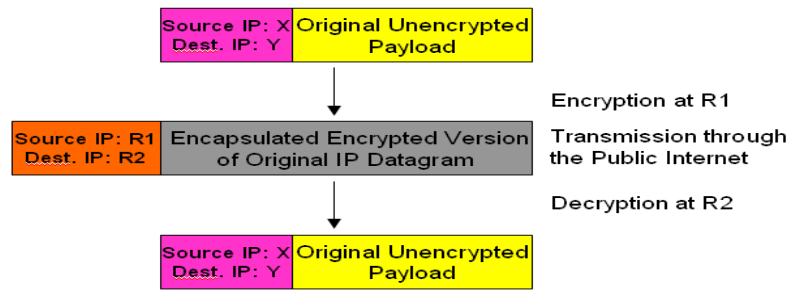

Figure 8: Structure of an IP Datagram during Different Phases of IP-in-IP Tunneling

The virtual private network (VPN) technology uses IP-in-IP tunneling [14] to encrypt and encapsulate the IP datagram that has the private IP addresses of the two end hosts with another IP header that has the source and destination IP addresses as the public IP address of the gateway routers for these two private networks. Each organization is required to have one or more gateway routers with a public IP address in order to facilitate communication over the public Internet. As the original IP datagram is encrypted, no intermediate forwarding host in the public Internet can look at the contents of the message. Figure 7 illustrates the notion of a VPN and Figure 8 displays the structure of an IP datagram as it goes through the different phases of IP-in-IP tunneling.

### 3.3 Secure Shell (SSH)

Secure Shell (SSH) [15] is a network protocol that allows a user to securely interact with remote machines by establishing a secure channel for data exchange. SSH replaced TELNET [16] and other insecure remote shell programs that were used in the past to send information in plaintext, including passwords, to remote systems. SSH encrypts the information sent over the insecure Internet and thus provides both confidentiality and integrity of data. SSH operates over a sequence of three phases as illustrated by the timeline diagram shown in Figure 9. The three phases are described below:

*Step* 1: *Host Identification* – The client machine needs to ensure that it is communicating with the remote machine it has been asked to by the application program, and not with another machine that is spoofing it. The server machine on the remote side also has the option to ensure that the user is connecting from the machine as it appears to be, and not from another machine that is spoofing it. This step is accomplished as outlined below:
- The client contacts the server and requests for its public-key certificate.
- The client maintains a list of public keys for server machines available to it. If it is asked to contact a machine for which it does not have a public key locally held, it will warn the user with a message telling that the public key reported by the server is not in the list of known hosts and ask the user whether the user wants to continue connecting.
- If the user agrees to continue connecting, the client verifies the authenticity of the Certifying Authority (CA) that issued the public key certificate for the server and if satisfied, accepts the public keys. The machine then adds the server's public keys to its personal list of host public keys.
- When the administrator has included the public key for the client machine in the per-machine list of known host public keys on the server machine, the server may want the client machine to prove that it is what it claims to be.
    - The server will create a "challenge" encrypted with the client's host public key and send it to the client. Only a genuine client machine will be able to decrypt this message with its private key. The client then sends the same challenge encrypted with the public key of the server. If the server when decrypting the message gets the same challenge it sent, the client is genuine.

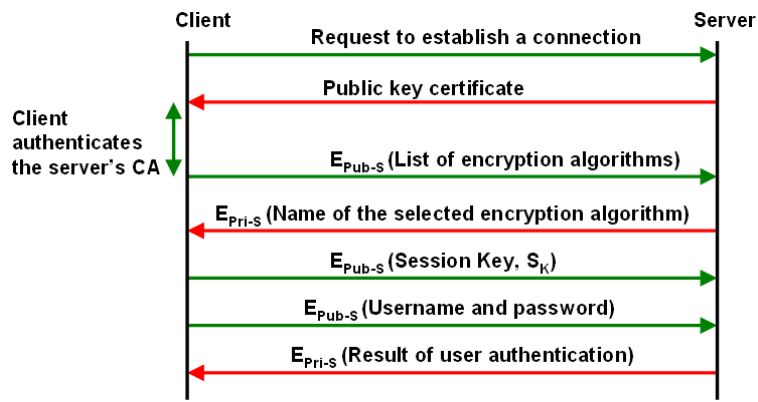

Figure 9: Steps to Establish a Secure Shell (SSH) Connection

*Step* 2: *Encryption* – The objective of this step is to establish a secure end-to-end link that supports encryption of the data transferred. Even the password and other authentication information are encrypted and are not transmitted in plaintext. This step is accomplished as outlined below:
- Once the host identification step is successfully done, the client sends a list of encryption algorithms it could use and their corresponding keys. This is sent encrypted with the public key of the server.
- The server decrypts the list with its private key and chooses the strongest encryption algorithm that it could handle from the list sent by the client.
- The server then notifies the selected encryption algorithm to the client by encrypting the notification using its private key.
- The client generates the appropriate secret session key for the encryption algorithm selected and notifies the session key to the server by encrypting the notification with the public key of the server.
- The server decrypts the notification with its private key and extracts the secret session key.

*Step* 3: *User Authentication* – In this step, the user proves to the server that he/she has the right to perform operations as a particular user on the server machine. This is accomplished as outlined below:
- The client asks for the username and password from the user, encrypts them with the server's public key and sends to the server.
- The server checks the validity of the username and password and if everything is fine, accepts the connection request by sending the confirmation encrypted with its private key.
- The client decrypts the confirmation with the server's public key and the client and server are all set to exchange data securely using the encryption algorithm selected and the secret session key agreed.

### 3.4 Transport Layer Security (TLS)

Transport Layer Security (TLS) [17] is the successor of the Secure Sockets Layer (SSL) [18] cryptographic protocol and it provides secure communication of the datagrams of the transport layer protocols as part of an end-to-end connection across the network. TLS has been used for a wide-variety of applications like web browsing, electronic mail, voice-over-IP, instant messaging and etc.

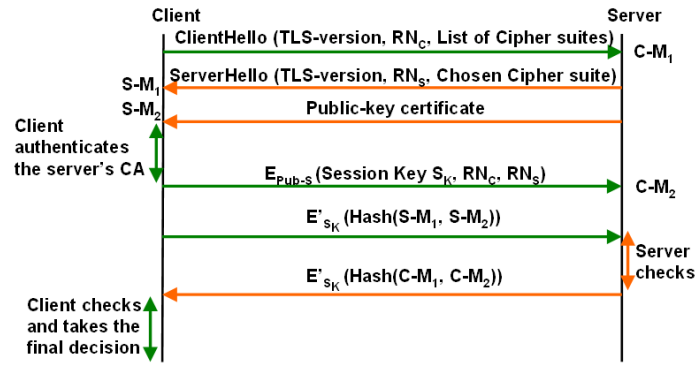

Figure 10: TLS Connection Establishment Mechanism

We now explain the sequence of steps to be followed to establish a TLS connection between a client and a server and it is pictorially illustrated in Figure 10:
- The client initiates the connection request by sending a *ClientHello* message to the server. This message has the following information: (i) The latest TLS-version supported by the client; (ii) A random number arbitrarily chosen by the client and (iii) A list of suggested cipher suites (i.e., the encryption algorithms to be used, the key exchange and authentication algorithms, as well as the hashing algorithms to generate message authentication codes).
- The server responds back with a *ServerHello* message that includes the following information: (i) The TLS version chosen by the server based on the version information submitted by the client; (ii) A random number arbitrarily chosen by the server and (iii) The cipher-suite chosen from the list of choices offered by the client.
- The server also sends its public-key certificate to the client. The client may contact the CA that issued the certificate and confirm that the certificate is authentic before proceeding. The server also has the option of asking for the client's public-key certificate by sending a *CertificateRequest* message, so that the connection can be mutually authenticated.
- The client generates a shared session key and sends it along with the client-side and server-side random numbers, all encrypted with the public key of the server. The client-side and server-side random numbers are merely sent to enhance each other's authentication.
- The server decrypts the message received with its private key and extracts the shared session key.

- The client then computes a hash of the messages received so far from the server using the hashing algorithm agreed upon, encrypts the hash value with the shared session key using the encryption algorithm selected and sends it to the server.
- The server decrypts the client's message with the shared session key using the decryption algorithm selected. The server then independently calculates the hash of all its messages to the client using the hashing algorithm agreed upon. If the hash value matches with the hash value in the message sent by the client, the server basically accepts the connection request from its direction. The server then computes a hash of all the messages it has received so far from the client and then sends it to the client by encrypting the hash value with the shared session key.
- The client decrypts the message with the shared session key and independently computes a hash of all the messages it has sent to the server. If the hash value locally computed matches with the hash value sent by the server, then the client has basically authenticated the server. Thus, a TLS connection is established.

### 3.5 IP Security

The IP Security Protocol suite (IPSec) [19] is implemented at the IP layer, so it does not require any change to existing transport layer and application layer protocols. IPSec is primarily designed to address the fundamental shortcomings of the IP layer such as IP address spoofing, wiretapping and session hijacking. The following two protocols are used to provide packet-level security for both IPv4 and IPv6:
◊ IP Authentication Header, AH (Next Header protocol ID: 51) [20] provides integrity, authentication and non-repudiation
◊ IP Encapsulating Security Payload, ESP (Next Header protocol ID: 50) [21] provides confidentiality, along with authentication and integrity protection.

#### 3.5.1 Security Association

The basis of IPSec is a Security Association (SA) [22], characterized by the set of security parameters agreed upon for a secure communication channel between two communicating hosts. Each host can have several SAs in effect for communication with different remote hosts. A SA is identified using a Security Parameter Index (SPI) – a 32-bit identifier and the IP address of the partner host on the other side of the SA. The SPI and the partner IP address are used to index to the Security Association Database (SADB) that has information about the characteristics of different SAs. A SA is characterized by the following parameters: Encryption algorithm, Encryption key, Encryption parameters like the Initialization Vector, Integrity/Authentication algorithms (keyed-HMAC algorithms [23] and the key) and Lifespan of the SA.

A SA is uni-directional. For two hosts to communicate in either direction, SAs have to be established separately in both directions. For host A to securely send data packets to host B, and make host B to believe that the data packet did come from host A, it should establish a SA with host B. Such a SA is said to be "outbound" at A and "inbound" at B. An IPSec header of a datagram sent from host A to host B, should have the secure features of the SA that is "inbound" at B and similarly the IPSec header of a datagram sent from host B to host A should have the secure features of the SA that is "inbound" at A.

Prior to establishing an IPSec SA, the two end hosts need to exchange their public-key certificates digitally certified by a trusted third-party certificate authority (CA). This is done through the Internet Key Exchange (IKE) protocol [24]. Once the two hosts have exchanged each other's public-key certificates, then they are said to have established an IKE Security Association (IKE SA). Establishing an IKE SA is a pre-requisite to establish an IPSec SA. The procedure to establish an IPSec SA is explained as follows:
- Host A wishing to send data packets to host B needs to establish an "inbound SA" with host B.
- Host A picks a SPI that has not been yet chosen for communication with B and sends a "SA Establishment Request" message to B which contains the following:
  - SPI for the inbound SA channel at host A (i.e., the outbound SA channel at host B)

- o Lifespan of the association – negotiable by host B
- o The packet-level security protocol chosen (AH or ESP) – negotiable by host B
  - ◊ If AH is chosen, then the list of keyed-HMAC algorithms that could be used is specified. Host B will choose one from this list if it wishes to receive packets from host A.
  - ◊ If ESP is chosen, then the list of keyed-HMAC algorithms along with the list of encryption algorithms and key-derivation functions that could be used will be sent.
- All negotiation messages (including the SA Establishment Request) are encrypted at the sender side using the receiver's public key and decrypted with the receiver's private key at the receiver side.
- Hosts A and B agree on a shared session key using the Diffie-Hellman exchange algorithm [25].
- The shared session key would be used for the keyed-HMAC algorithm.
- Each host uses the shared session key and the key-derivation function agreed upon to derive the secret key to be used for encryption and decryption of the data at hosts A and B respectively.

### 3.5.2 Authentication Header (AH)

AH provides integrity and data origin authentication for IP datagrams. AH operates on the top of IP, using the IP protocol number 51. The different fields in an AH are described below (also refer Figure 11). The structure of an original IPv4 datagram and IPv4 datagram with AH is shown in Figure 12.

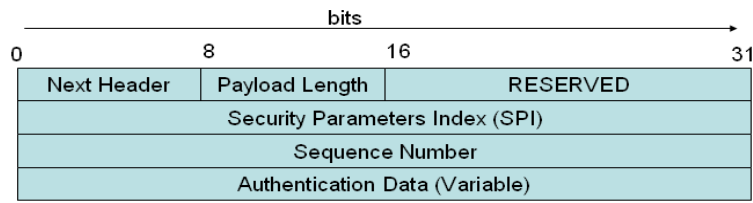

Figure 11: Structure of an Authentication Header (AH)

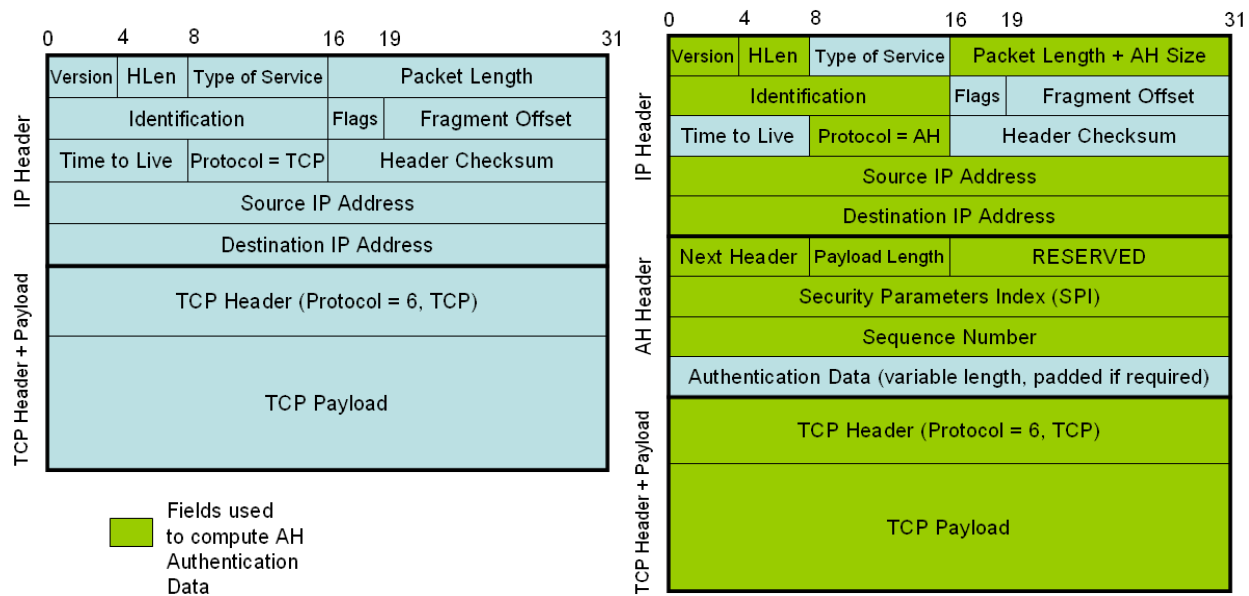

Figure 12: Original IPv4 Datagram and IPv4 Datagram with AH Header

- *Next Header*: Identifies the transport layer protocol
- *Payload Length* (*AH Length*): Indicates the length of the whole AH in 32-bit words
- *Reserved*: Indicates that this field is reserved for future use and it must be set to zero

- *SPI*: Identifies the security association
- *Sequence Number*: Identifies the datagrams sent as part of a SA. This field is a monotonically increasing identifier and is used to assist in anti-replay protection
- *Authentication Data*: Contains the integrity/authentication check value (keyed-HMAC) calculated over the entire packet, including the header fields that do not change at the intermediate hosts. The size of the keyed-HMAC may vary with each SA and may not be exactly multiple of 32 bits. If this is the case, the HMAC will be padded.

### 3.5.3   Encapsulated Security Payload (ESP)

ESP provides origin authentication, integrity and confidentiality protection for the IP datagrams. The different fields in an ESP header are described below (also refer Figure 13). The structure of an original IPv4 datagram and IPv4 datagram with ESP header is shown in Figure 14.

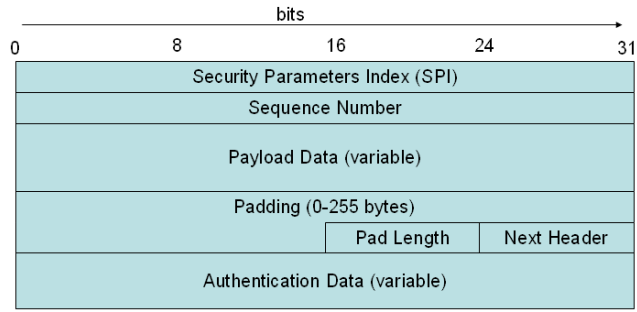

Figure 13: Structure of an Encapsulated Security Payload (ESP) Header

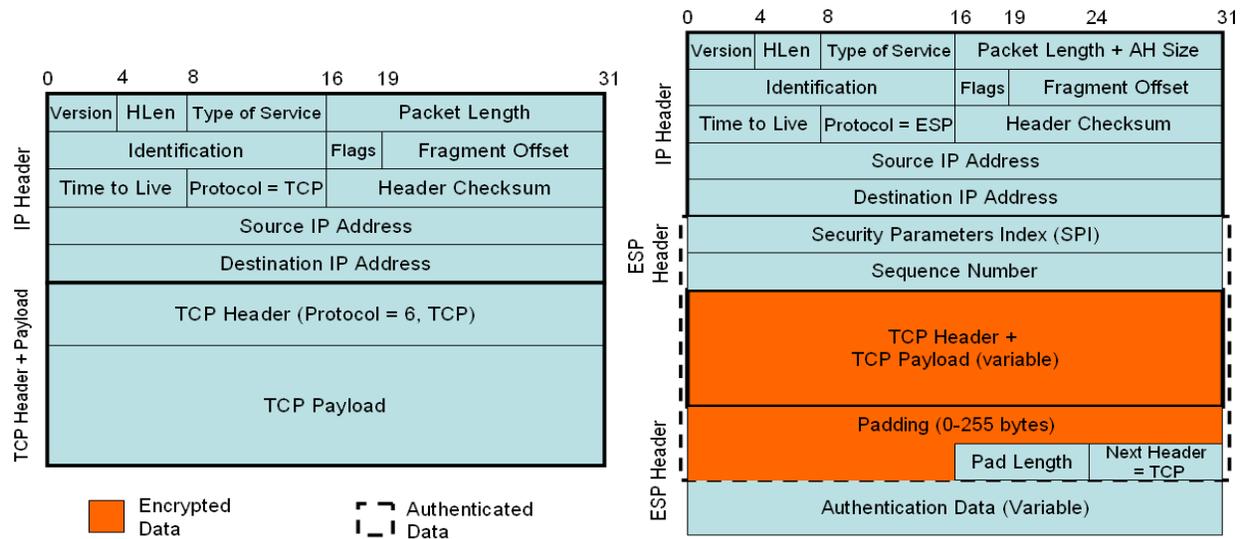

Figure 14: Original IPv4 Datagram and IPv4 Datagram with ESP Header

- *SPI*: Identifies the security association
- *Sequence Number*: Identifies the datagrams sent as part of a SA. This field is a monotonically increasing identifier and is used to assist in anti-replay protection
- *Payload data*: Indicates the data to be transferred
- *Padding*: Used with certain block ciphers for padding the payload data to a full block length.
- *Pad length*: Indicates the size of the padding in bytes
- *Next Header*: Identifies the transport layer protocol

- *Authentication Data*: This is the integrity/ authentication check value (keyed-HMAC) calculated over only the SPI, Sequence Number in the ESP header, the actual data, padding data, pad length and the Next Header field.

### 3.6 Kerberos

Kerberos [26] is an authentication protocol used by processes/hosts communicating over an insecure network to verify each other's identity in a secure manner. It is based on the idea that a central server provides authenticated tokens called "tickets" to requesting applications. A ticket is an unforgeable, non-replayable, authenticated object. The security of the protocol depends on the assumption that the participating machines maintain loosely synchronized time. The four entities involved in Kerberos are: (i) Authentication Server, AS; (ii) Ticket Granting Server, TGS; (iii) Service Server; SS and (iv) Ticket Granting Ticket, TGT. A client authenticates itself to the AS once and obtains a ticket that can be used to obtain additional tickets from the SS without requiring the client to re-authenticate itself for every service requested. The sequence of steps of the protocol is described below (also shown in Figure 15):

#### 3.6.1 Kerberos Protocol Steps

*Step* 1: *User Client-based Logon* – The user submits the username and password information to the client machine. The client machine uses a one-way function on the entered password to compute the secret key for the user.

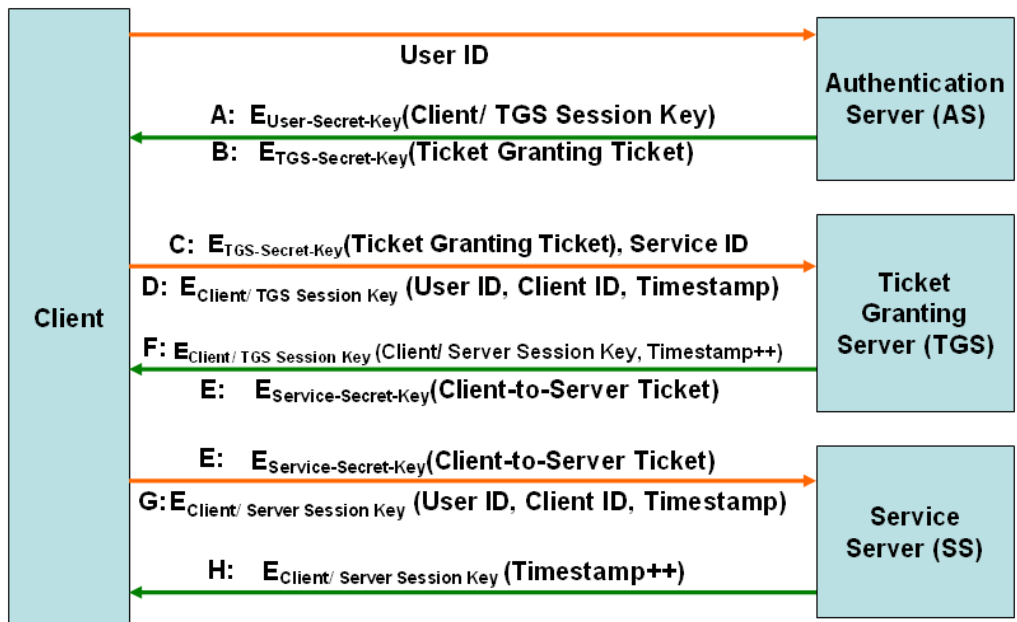

Figure 15: Kerberos Protocol Steps

*Step* 2: *Client Authentication* – The client sends the username in plaintext to the Kerberos AS. The AS checks the username in its database, and if an entry exists, the AS sends back two messages:
- Message A: contains the Client/ TGS session key encrypted with the secret key for the user (derived from the user's password at the AS).
- Message B: contains the Ticket-Granting Ticket, TGT, which includes the following: (i) username, (ii) network address of the user's client machine, (iii) validity period of the TGT and (iv) Client/ TGS session key. The TGT is encrypted using the secret key of the TGS.

Once the client receives messages A and B, the client decrypts message A with the secret key of the user and extracts the Client/ TGS session key.

*Step* 3: *Client Service Authorization* – The client sends the following two messages to the TGS:
- Message C: contains the TGT from message B and identification for the requested service.
- Message D: contains the authentication information for the user/client. The authentication information submitted includes the username, the network address of the user/client machine and a timestamp. All of this information is encrypted using the Client/ TGS session key.

After receiving message C, the TGS decrypts the message with its secret key and extracts the Client/TGS session key. The TGS then decrypts message D using the Client/ TGS session key and sends back the following two messages to the client:

- Message E: contains the Client-to-Server Ticket, which includes the username, the network address of the user/client machine, the validity period the ticket and the Client/Server session key. The ticket is encrypted with the secret key of the server for the service requested.
- Message F: contains the Client/Server session key and the timestamp of message D incremented by 1. Both the key and the timestamp are encrypted with the Client/ TGS session key.

After the client receives messages E and F, it uses the Client/ TGS session key to decrypt message F to extract the Client/ Server session key.

*Step* 4: *Client Service Request* – The client sends the following two messages to the SS: (i) message E (received from the TGS) and (ii) message G, containing the username, the network address of the user/client machine and the timestamp, all encrypted with the Client/Server session key. The SS decrypts message E using its secret key and extracts the Client-to-Server ticket and the Client/Server session key. The SS decrypts message G using the Client/Server session key and extracts the user/client identification information. If the user/client identification information in message G matches with the user/client information in the Client-to-Server ticket, the SS increments the timestamp information in message G by 1. The incremented timestamp is encrypted using the Client/Server session key and sent back to the client (as message H). The client on receiving message H decrypts the message with the Client/Server session key. If the timestamp value in message H is the value expected by the client, the client trusts the server and start sending service requests to it.

### 3.6.2   Kerberos Advantages

- A user's password is not sent on the wire (either in plaintext or ciphertext) during session initiation.
- Kerberos provides cryptographic protection against spoofing. Each service access request is mediated by the TGS, which knows that the identity of the user/client is authenticated by the Kerberos AS and processes the user/client request encrypted with the Client/TGS session key
- As each ticket has a limited validity period, long-term cryptanalytic attacks cannot be launched.
- Kerberos assumes that the clocks across all the clients and the servers are synchronized. A host responds back only if the request messages have timestamp value close to the current time at the host.
- Kerberos provides mutual authentication. The TGS and SS can respectively get access to the Client/TGS session key and the Client/Server session key only after they can decrypt the messages containing these keys with their appropriate secret keys. The client uses this approach to indirectly authenticate the servers.

### 3.6.3   Kerberos Weaknesses

- Kerberos requires continuous availability of a trusted ticket-granting server for all access control and authentication checks.

- Authenticity of servers requires a trusted relationship between the TGS and every service server.
- Timely transactions are required to reduce chances of a user with genuine ticket being denied service.
- Password guessing could still work to get the valid secret key for a user. The whole system is still dependent on the user password.
- Kerberos does not scale well as the number of service servers is increased. The TGS has to maintain a trustworthy relationship and maintain the secret key for each SS. Adding backup service servers further complicates the situation.
- Network services cannot be accessed without obtaining Kerberos authentication. All applications run by the users in the network need to go through Kerberos authentication.

## 3.7 Firewalls

A firewall is a device that filters traffic between a "less trustworthy" outside network and a "protected" inside network [6]. A firewall is often implemented in software and is basically a code running on a dedicated computer located in the periphery of the network to be protected. All traffic exiting and entering the network should go through the firewall. Hence, the firewall is often a kind of bottleneck for network performance. This is also the reason why non-firewall functions are not normally performed at a computer running the firewall and hence it becomes hard for an attacker to get access to the firewall code by making use of the program vulnerabilities in the non-firewall code, if any exists.

Firewall design often makes use of either "default-deny" or "default-allow" approach [6]. In the default-deny approach, the firewall will be configured to allow only explicitly stated network connections. Any network traffic that does not match with the firewall ruleset will be denied access. In the default-allow approach, a firewall will be configured with the set of rules that describe what network traffic should be specifically blocked. Network traffic that does not match with the firewall ruleset will be allowed access. The security policies implemented at a firewall depend on the specific threats against which the inside network needs to be protected. Firewalls can be classified into four different categories: Packet filters (ii) Stateful inspection firewall (iii) Application proxy firewall and (iv) Personal firewall

### 3.7.1   Packet Filters

A packet filtering firewall [6] controls access to packets based on the network address (representing the network of the source or the destination) or IP address (source or destination IP address) or port numbers (representing specific transport layer protocols). Packet filter firewalls can be of two types: (i) Egress filter – packets will be sent out (or not sent out) only to specific networks and/or belonging to specific transport layer protocols; (ii) Ingress filter – packets belonging to (or not belonging to) certain specific networks or specific transport layer protocols could be let in. In order to combat IP spoofing, packet filter firewalls are often configured not to let inside a packet that has an IP address corresponding to the internal network being protected. As we increase the number of specific networks, IP addresses and transport layer protocols whose traffic needs to be blocked, the code for a packet filter firewall becomes lengthy.

### 3.7.2   Stateful Inspection Firewall

A packet filter firewall is stateless: it operates on packets on an individual basis and does not store or apply the state information pertaining to the action taken on packets processed earlier. Stateful firewalls examine each packet with regards to their overall placement in the packet series belonging to a specific connection. A stateful firewall [6] maintains records of all connections passing through it and will be able to determine whether a packet is part of any existing connection or a new connection. The connection state will be used to trigger specific rules for the firewall. Some of these could be: (i) The firewall can stop packets belonging to an already terminated TCP connection from getting inside the network or to a TCP connection that has been not yet established. (ii) The firewall may not let more than a specific amount of data to be transferred from the protected network to a specific network or an outside IP address. (iii) The firewall may not allow beyond a maximum number of TCP connections per IP address.

### 3.7.3 Application Proxy Firewall

The packet filters and the stateful inspection firewalls take actions by only looking at the headers of the data packets. An application proxy firewall acts as an intermediate gateway attempting to look not only at the packet headers but also at the data inside the packets entering or leaving the network to be protected. All communication between the users of the internal protected network and the Internet has to pass through the proxy server instead of allowing users to directly communicate with servers on the Internet.

An application proxy firewall [6] is dual homed (refer Figure 16): for client applications running inside the protected network, the application proxy firewall acts as a proxy server and for server applications in the outside network, the application proxy firewall acts as a proxy client. The request generated by an internal user (client) to connect to an external service goes through the application proxy firewall that runs a proxy server for the particular service being requested. The proxy server evaluates the connection request and decides to permit or deny the request based on a set of rules that are managed for the individual network service. Only packets that comply with the services of the application protocols are allowed by the proxy server, which forwards the accepted request to the proxy client. The proxy client then contacts the real server in the Internet on behalf of the real client in the protected network and proceeds to relay requests from the proxy server to the real server and the responses from the real server to the proxy server. The proxy server relays requests and responses between the proxy client and the real client in the protected network.

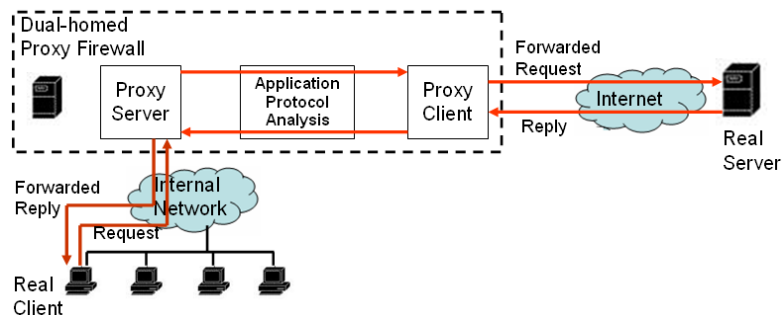

Figure 16: Application Proxy Firewall

The application firewall could also act as the proxy server for a real client in the outside network and as a proxy client for a real server inside the protected network. In this case, the real client contacts the proxy server, which evaluates the request and forwards the request to the proxy client. The proxy client contacts the real server running in the internal network and forwards the response from the real server to the proxy server. The proxy server forwards the response to the real client in the outside network.

### 3.7.4 Personal Firewall

Personal firewalls [6] are very much needed for standalone machines connected to the Internet through various means like dial-ups, cable modems or DSL connections. Having a separate firewall computer to protect a single computer system can be too expensive and complex. A personal firewall is an application program running on a specific computer system. The firewall screens the incoming and outgoing traffic for the computer system and blocks the unwanted traffic from entering or leaving the system. The user could configure the personal firewall to accept traffic only from certain sites or not from specific sites and to generate logs of the past activities. The personal firewall can also be configured to function as a virus scanner so that any incoming data to the system will be first scanned for any potential virus infection.

### 3.7.5 What Firewalls Can and Cannot Block

A firewall can be held responsible if and only if it is the only means for traffic to leave or enter the inside network being protected. If the inside network has a host that is connected to the outside network through a modem, then the whole of the inside network is exposed to the outside network through the modem and the host. The firewall cannot be held responsible if certain malicious traffic manages to get into the network through the modem and the host [6]. A firewall is often a single point of failure for the network being protected. Modern day networks have layered firewall architecture (refer Figure 17) comprising multiple firewalls: a screening router implemented with the packet filter, followed by a proxy firewall and then followed by a personal firewall at every host in the network. The security policies of a firewall must be constantly updated to take into account the latest intrusion attempts and potentially harmful application software that come into existence.

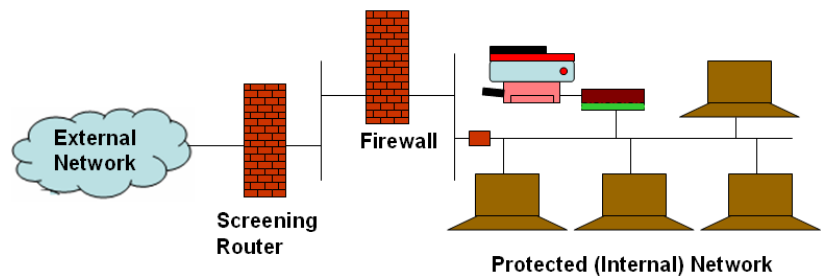

Figure 17: A Layered Firewall Architecture

## 3.8 Secure E-mail

Electronic mail (e-mail) has become a common communication method for both business and ordinary users. An e-mail, while propagating in the network channels, is very public and is exposed in plaintext at every point from the sender's system to the recipient's screen. This section analyzes the key requirements for a secure e-mail and explores two designs that satisfy one or more of these key requirements.

### 3.8.1   Key Requirements for Secure E-mail

Any design for secure e-mail should take into consideration that the protection measures should be enforced within the body of the message as the existing e-mail network in the Internet should not be changed in order to provide e-mail security. The key requirements [6] for secure e-mail are as follows:
- Confidentiality: E-mail contents should not be exposed on the path from the sender to the receiver.
- Integrity: The receiver should see in the e-mail, the same content which the sender sent.
- Authenticity: Receiver should be able to verify that the e-mail message indeed came from the sender.
- Non-repudiation: The sender of the e-mail cannot deny having sent the message.

### 3.8.2   Secure E-mail Design for Confidentiality, Non-Repudiation and Sender Authenticity

One could provide confidentiality by encrypting the message (before transmission) so that the message gets transmitted in ciphertext in the network channels and can be seen in plaintext only at the receiver after a successful decryption. In addition to ensuring confidentiality during transmission, it is essential for the receiver to verify the authenticity of the sender and also use the e-mail as a proof that the sender did send the message and cannot deny sending such a message. A plausible design to achieve all of this is explained below and it is also pictorially represented in Figure 18. This is also the basic idea behind the Pretty Good Privacy (PGP) secure e-mail software [27] used for sending text messages in secured fashion.

The payload portion of the e-mail contains the encrypted version of the sender's public key certificate and the encrypted version of the original e-mail header and the e-mail message. The sender's public key certificate is encrypted with the receiver's public key and included in the payload portion. The original message and the e-mail header are first encrypted with the sender's private key (provides non-repudiation and sender authenticity) and the ciphertext resulting from this encryption is further encrypted with the receiver's public key (provides confidentiality) and the resulting ciphertext is included in the payload portion of the e-mail transmitted in the network channels. The header of this e-mail is a plaintext copy of the e-mail header enclosed in the encrypted format in the payload portion.

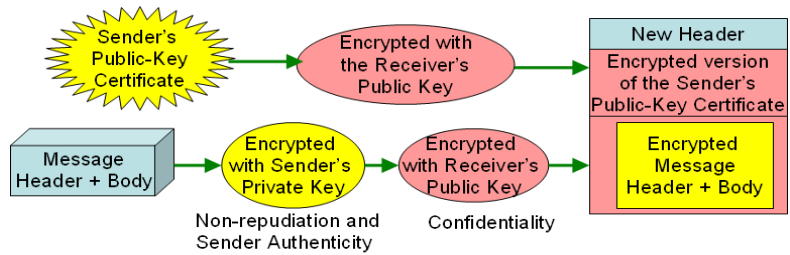

Figure 18: Secure E-mail Design for Confidentiality, Non-Repudiation and Sender Authenticity

### 3.8.3 Secure E-mail Design for S/MIME

S/MIME (Secure/ Multi-purpose Internet Mail Extensions) [28] is a secure e-mail standard commonly used in the Internet. It satisfies all the four requirements for secure e-mail design. The basic idea behind the design of S/MIME is pictorially represented in Figure 19 and is discussed below.

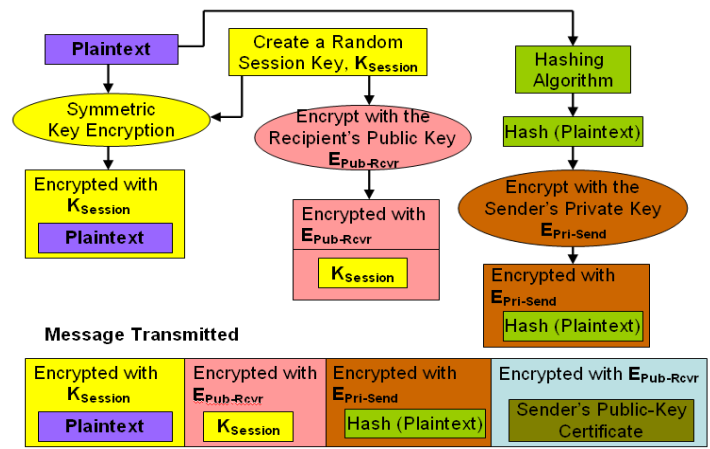

Figure 19: Secure E-mail Design for S/MIME

The payload of the e-mail sent from the sender to a receiver has four components:

(1) *The encrypted version of the original plaintext (e-mail message header and body)*: A random session key is generated at the sender side and is used as the secret key for this symmetric encryption. This component satisfies the requirement of providing confidentiality.

(2) *The encrypted version of the random session key used for the symmetric encryption of the original message header and body*: The session key is encrypted with the receiver's public key. This approach of generating a random session key to encrypt every message sent and also sending the session key along with the message avoids the offline use of a key distribution algorithm [25] between the sender and the receiver. This component satisfies the requirement of providing confidentiality.

(3) *The encrypted version of the hash of the original e-mail message header and body*: A hash value of the original e-mail message header and body is computed using a standard hashing algorithm (like SHA1 [29]). This hash value is encrypted with the private key of the sender. This component satisfies the requirement of providing message integrity.

(4) *The encrypted version of the public-key certificate of the sender*: The public-key certificate of the sender is encrypted with the public key of receiver so that it can be decrypted only by the receiver and used for extracting the hash value of the plaintext message header and body. This component satisfies the requirement of providing sender authenticity and non-repudiation.

## 4 Conclusions

The crux behind network security is to ensure access to the network and its data for authorized hosts/users and deny access to unauthorized hosts/users. A secure network needs to have tamper-proof communication media and resilient protocol mechanisms that can avoid or reduce the chances of an attack. A close look at the classical network attacks described in Section 2 reveals that IP spoofing has been behind the success of most of these attacks. Hence, it has become a design requirement that in addition to authenticating the application users, it is also essential to authenticate the networks and hosts from which the application users are communicating in the Internet. Protocol mechanisms like SSH, TLS, IPSec and Kerberos ensure that the above requirement is being taken care of and reduce the chances of spoofing-based attacks. A single security control mechanism cannot combat all kinds of network attacks. The security control mechanism(s) chosen for a network should be based on the specific threats that currently exist for the network. There is always a tradeoff between using security control mechanisms as mere plug-in modules and making them more embedded with the core functionality of the protocols in the TCP/IP stack. It would be better for a security control mechanism to require changes to be made only in one particular layer of the Internet protocol stack rather than all the layers.